**Fuzzy data: XML may handle it**

Keywords: Fuzzy data, Model flexibility, XML schema

Author: Ralf Schweiger; Simon Hoelzer MD, Joachim Dudeck MD

Institute of Medical Informatics, Justus-Liebig-University Giessen, Germany

_________________________________________

Author Contact Address:

Ralf Schweiger
Heinrich-Buff-Ring 44
35392 Giessen
Germany

tel.:+49-641-9941370
fax.:+49-641-9941359
mailto:Ralf.Schweiger@informatik.med.uni-giessen.de

**Fuzzy data: XML may handle it**

***Abstract***

Data modeling is one of the most difficult tasks in application engineering. The quality of the data model often determines the quality of the application itself. As a consequence, data modeling has to be done very carefully. The engineer must be aware of the use cases and the required application services and at a certain point of time he has to fix the data model which forms the base for the application services. However, once the data model has been fixed it is difficult to consider changing needs. This might be a problem in specific domains, which are as dynamic as the healthcare domain. With fuzzy data we address all those data that are difficult to organize in a single database. In this paper we discuss a gradual and pragmatic approach that uses the XML technology to conquer more model flexibility. XML may provide the clue between unstructured text data and structured database solutions and shift the paradigm from „organizing the data along a given model" towards „organizing the data along user requirements".

*Introduction*

Most heavily, XML is discussed as a message format (1,2). Existing interchange formats are often limited with respect to the number of available delimiters and with respect to the number of hierarchical levels. With XML on the other hand we may easily invent new delimiters and the simple start-end-delimitation of XML can express any hierarchical structure. In addition, an arriving XML message can be easily parsed and transformed using the DOM and XSL facility. XML consequently provides the communication framework that will facilitate the development of communication interfaces (3).

However, we might use XML also as a storage format. Database solutions are well accepted in domains where the data structures are approximately clear. Clear in this context implies a „broad consensus" and the „validity over a long period of time". However, this is not true for all domains. Especially in the healthcare domain we often find narrative documents that are difficult to exploit with respect to their content. Experts in medical information science consequently postulate the insertion of more structure and more codes into medical documents (4). The reasons for this situation are manifold. Healthcare data are document oriented, that means the healthcare data are organized into cohesive communication units that contain different kinds (administrative and clinical) of data. Such data are difficult to organize in a single database. Another reason for this situation is the fact that healthcare data need to be very flexible with respect to their structure. It must be possible to put a comment into any context of the clinical document. Any model that restricts the proper, i.e. the context related accommodation of the information would fail sooner or later. A model must not put constraints on the documentation process, but has to obey the user requirements. Finally, we cannot reduce clinical documents to a set of related data items since items often have a descriptive context. We consequently need the possibility to freely mix up data items with narrative text. Database schemas are often not flexible enough to satisfy these constraints (cohesive form, flexibility to change and narrative context).

With XML on the other hand, we may invent markup for the identification of single items in textual documents, i.e. we add structure to the data rather than adding data to a given structure. The data consequently remain in a cohesive form. In addition, the structure remains flexible since we can easily add new markup, i.e. structure to the documents. Besides, XML provides the concept of mixed content models allowing to freely mix up single items with narrative text. The problem of the XML approach reveals, when we have to manage a large set of related documents, e.g. the reports of an entire pathology institute. We run the risk of losing control over the items contained in the various documents. What we need is a sort of data model or schema that can be used to establish search strategies upon the document base. The XML schema approach (5) may help to solve the problem.

### The concept

The idea is to let the document author extend the document where and when necessary and to automatically update an XML schema definition which represents the overall document model. The XML schema definition keeps track of the items contained in a set of related documents and can be used to establish search strategies upon the document set. XML consequently provides the means to preserve the flexibility of free text documents AND to schematize the data at the same time. XML may provide the clue between unstructured text data and structured database solutions and shift the paradigm from organizing the data along a given and often fixed schema towards organizing the data along changing user requirements.

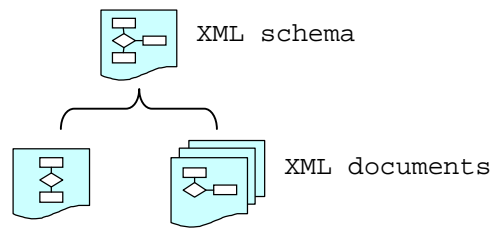

**Figure 1**: The XML schema as a superset and summary of the items contained in a set of related documents. We may change the schema as easily as the documents (both are XML files) thus combining the flexibility of free text with the schematization of a database.

## *Method*

The following figure outlines the overall architecture of our approach. In the center of the architecture is a software component, which will be referred to as manager. The manager uses an XML schema definition of the document model to automatically generate and organize a corresponding user front and storage back end. As a consequence, the application system can be easily adjusted to changing user requirements by simply providing a different model description. If needed, the manager will reorganize existing data. In addition, the manager changes the user interface so that the end user can enter data instances to the newly added model items. The main difference of our approach compared to a database approach is the fact that the meta data, i.e. the content model can be updated as easily as the data itself. In addition, the manager handles both, the user front and the storage back end. The model description, the user front end (data entry and data presentation) and the storage back end will be now described in separate sections.

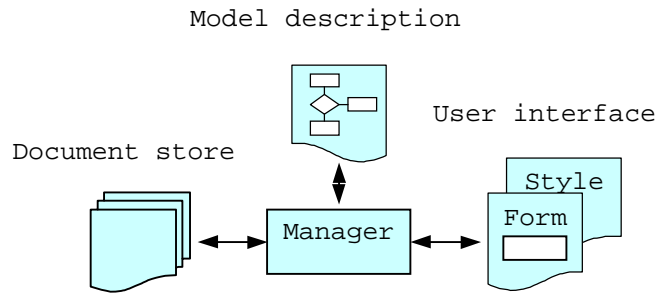

**Figure 2**: XML based application system that improves the model flexibility. The manager automatically adjusts the entire application system to the latest model description (XML schema definition). The user himself may change the document model in order to satisfy the latest documentation requirements.

## Model description

For the description of a document model, we use a standard XML schema definition (Schema). Figure 3 presents the XML schema of a simple clinical report. The root element of a corresponding model instance, i.e. an XML document that validates against the XML schema, is the report element. According to its type, the report element contains a report date, a sender and a receiver of the report, a patient that the report relates to, a report subject (reason of communication) and a report content (clinical findings etc.). The sender and the receiver of the report are both of type PersonType, i.e. they have a name and an address. The PatientType is derived from the PersonType by extension that means the patient has a name, an address and in addition a date of birth and a gender. Elements without an explicit type such as the report date and the report subject are assumed to have the simple type „string". The minOccurs and maxOccurs attributes of the XML schema namespace are used to define optional elements such as the person address and repeatable elements such as the address telecommunication (telephone, facsimile, email etc.).

The standard XML schema markup (default namespace) is mixed up with the markup of a nonstandard user namespace. The user namespace defines a few XML attributes such as prompt and form which allow the schema designer to easily assign and relate elements of the user form to elements of the document model. The usage of a nonstandard namespace for the description of data forms is just pragmatism and might be replaced by any other namespace with similar expressiveness. All elements in the XML schema with simple types correspond to data entry forms in the user interface. The gender element e.g. has the prompt „Geschlecht" and the possible gender values m(ale), f(emale) and no value are represented by the prompts „mann", „frau" and „undefined" which are offered to the end user in a selection box. The report content element has the prompt „Befund" and corresponds to a text area with eleven rows. Default values for the prompt (element name) and the form (single line of text) attribute further simplify the schema definition.

```
<schema
    xmlns="XML schema namespace"
    xmlns:user="namespace for the description of the user interface">

<element name="Report" type="ReportType" user:prompt="Bericht"/>

<complexType name="ReportType">
<element name="Date"/>
<element name="Sender" type="PersonType" user:prompt="Von"/>
<element name="Receiver" type="PersonType" user:prompt="An"/>
<element name="Patient" type="PatientType"/>
<element name="Subject" user:prompt="Betreff" user:form="narrate 4"/>
<element name="Content" user:prompt="Befund" user:form="narrate 11"/>
</complexType>

<complexType name="PersonType">
<element name="Name"/>
<element name="Address" type="AddressType" minOccurs="0"/>
</complexType>

<complexType name="PatientType" base="PersonType" derivedBy="extension">
<element name="DoB" user:prompt="Gebdat"/>
<element name="Gender" user:prompt="Geschlecht" user:form="mann/m,frau/f"/>
</complexType>

<complexType name="AddressType">
<element name="Institution" user:form="narrate 2"/>
<element name="Street" user:prompt="Strasse"/>
<element name="City" user:prompt="Ort"/>
<element name="Telecommunication" maxOccurs="unbounded"
user:prompt="Telekomm"/>
</complexType>

</schema>
```

**Figure 3**: XML schema definition of a simple clinical report.

## User front end

Figure 4 and Figure 5 represent the user interface of the application system. The manager generates the user form from the XML schema definition in Figure 3 and enables the end user to enter instances of model elements, i.e. to assign data to the elements in the XML schema. When selecting an element instance (radio button), the applicable functions automatically appear on top of the data entry form. The functions of the user front end can be subdivided into editor functions and document management functions.

Editor functions may be used to open and close composed and narrative element instances thus controlling the document details. The content of closed element instances is presented to the user so that the user can immediately open the right instance in the case of many element instances. The newInstance (creation of a new element instance) and delInstance (deletion of an element instance) function are derived from the element occurrence defined in the XML schema. The selected content element e.g. is not repeatable (we would have to specify maxOccurs="unbounded" in the XML schema). As a consequence, the newInstance function does not appear. Beside of instance functions, the user interface may also offer model functions to the user. Model functions such as the newEl function enable the user to insert new elements into the document model.

Examples for document management functions are shown in Figure 4. The getModel function allows the user to select a different document model, e.g. a different version of the pathology report or even a different model domain such as laboratory. The newDoc function creates an empty document from the XML schema that can be used to enter either a completely new report or to enter only a few selecting data and to retrieve the matching documents by the getDoc function. The currently selected documents are listed in the left frame of the Web browser window and the user can choose a document from the list for update purposes. The styleDoc function allows the user to see the presentation style (Figure 5) of the data before he stores them in the document base using the putDoc function. The presentation style of the document model is defined in a stylesheet associated with the current content model.

Figure 4: User form for data entry.

**Figure 5**: Presentation of data.

## Storage back end

Figure 6 shows the back end representation of the data entered by the user. The data are logically stored as XML documents that are valid against the corresponding XML schema definition. The manager does not care about the physical details of the document store, which have been wrapped by some logical Application Programming Interface (API).

```
<?xml version="1.0" encoding="ISO-8859-1" ?>
- <Report>
    <Date>13. Apr 1999</Date>
+ <Sender>
+ <Receiver>
- <Patient>
    <Name>Ralf Schweiger</Name>
    <Address />
    <DoB>18.11.1968</DoB>
    <Gender />
  </Patient>
    <Subject>Einsendung vom 09.04.1999 Kennung 9907930 1 PE Antrum 2 PE
       Corpus</Subject>
    <Content>Es handelt sich um eine leichtgradige chronische Gastritis, ohne HLO
       (1,2). Im vorliegenden Material kein Anhalt für eine epitheloidzellig-
       granulomatöse Entzündung, kein Anhalt für einen Morbus Crohn und kein
       Anhalt für Spezifität oder Malignität.</Content>
  </Report>
```

**Figure 6**: XML structured data as the result of the data entry.

## *Discussion*

In this paper we present an XML based approach towards more flexible data models. The approach uses an XML schema definition for the description of a document model. Such an XML schema definition can be updated by the application engineer or even the user himself. The adjustment of the entire application system, i.e. the user front and the storage back end is managed automatically. As a consequence, new model requirements are easy to satisfy. With such an approach we may handle fuzzy data, i.e. data that are difficult to accommodate in a single database. In this section we will discuss the role of XML in our approach.

The ultimate problem of our approach can be phrased in the following question: How can we change the structure of single documents and track the structure (model) of a set of documents at the same time? Without an overall model we would not know what elements can be searched for in the document base. The answer is that the model needs to be as modifiable as the data itself. And this is exactly the point where XML may contribute to the solution of our problem. The standard XML schema language is well suited for the explicit description of document models. In addition, XML files, and this includes the model instances as well as the model itself, are easy to transform with standard XML facilities such as DOM and XSLT.

For the description of the document model we prefer an XML schema definition over a Document Type Definition (DTD). The XML schema approach supports more abstraction concepts thus allowing a higher reuse of definitions. This shall be illustrated at the PersonType definition. The PersonType defines an element type respectively an element class, which is reused for the sender and the receiver of a report. The XML schema consequently supports the abstraction concept of „classification" which can be already found in the DTD (Document Type Definition) approach. Furthermore, the XML schema supports a derivation concept that has no counterpart in the DTD. The PatientType e.g. is derived from the PersonType by extension, i.e. the PersonType is a generalization of the PatientType. Abstraction concepts such as classification and generalization allow a high reuse of definitions.

The model flexibility can be illustrated at the gender element of the presented XML schema definition (Figure 3). The gender element has been added to the schema in spite of the fact that existing documents would no longer comply with this new schema definition. The manager automatically extends the user form and lets the gender value undefined. The user can now define the gender value and save it in the document store. The key point in our approach is that the manager always uses the XML schema definition to organize the data in the application system.

XML may also facilitate the development of communication interfaces to other systems. This includes the export of documents to another system as well as the import of existing data into a given document. When sending a given document to another system, we usually have to transform the local storage format into a standard message format. XML supports such a format transformation. The Document Object Model (DOM) e.g. parses an XML document into its elements and provides methods for the ease of document navigation and document manipulation from within a programming environment (6). The eXtensible Styling Language (XSL) supports the transformation of existing documents into a different document style, e.g. a presentation style or a message style (7). The fact that XML is increasingly preferred as the interchange format for messages will further facilitate the development and the maintenance of communication interfaces.

Beside of the data export, the user may also retrieve existing data and integrate these data seamlessly into a given document. For this purpose, the user namespace may define a get attribute that can be added to any element in the XML schema and whose value is a standard URL. The URL defines the location of the remote data. For the integration of the patient data from an existing patient system e.g. we may add a „user:get" attribute to the patient element with the URL instance

http://patientHost/patientServer?service=getPatientData&argument=/Report/Patient/Name

The URL enables the manager to instantiate the arguments from the document content, to send the service value and the argument instances to the server and to integrate the returned data into the given document. The argument is defined using the standard XML Path language (8). The XPath "/Report/Patient/Name" e.g. instructs the manager to read the patient name from the document instance before contacting the remote server. The manager will complain if the patient name has not yet been entered. The data retrieval itself is controlled by the user using the getInstance function.

### Conclusion

Inflexible data models can lead to low acceptance of application systems in domains which are as dynamic as the healthcare domain. Our approach to this problem is to let the user requirements change the model description, i.e. to adjust the structure to the data instead of adjusting the data to the structure. The key concept of our solution is to make the structure, i.e. the XML schema definition as modifiable as the data itself. Necessary adjustments of the user front & storage back end are managed automatically. The XML technology provides a proper means in terms of XML Schema, XSLT, XPath and DOM to implement such an approach.